\documentclass{aa}  

\usepackage{graphicx}
\usepackage{comment}
\usepackage{txfonts}
\usepackage{soul}
\usepackage{color}

\usepackage{xcolor}
\usepackage{multirow}

\begin{document} 

   \title{Indication of a fast ejecta fragment in the atomic cloud interacting with the southwestern limb of SN 1006}

   \author{R. Giuffrida
          \inst{1,2}, M. Miceli\inst{1,2}, S. Ravikularaman\inst{3}, V. H. M. Phan \inst{4,5}, S. Gabici\inst{3}, P. Mertsch \inst{5}, S. Orlando\inst{2}, F. Bocchino\inst{2} 
          }

   \institute{Dipartimento di Fisica e Chimica E. Segr\`e, Universit\`a degli Studi di Palermo, Piazza del Parlamento 1, 90134, Palermo, Italy
         \and
             INAF-Osservatorio Astronomico di Palermo, Piazza del Parlamento 1, 90134, Palermo, Italy
        \and
            Université de Paris, CNRS, Astroparticule et Cosmologie, F-75013 Paris, France
        \and
            Sorbonne Université, Observatoire de Paris, PSL Research University, LERMA, CNRS UMR 8112, 75005 Paris, France
        \and    
            Institute for Theoretical Particle Physics and Cosmology (TTK), RWTH Aachen University, 52056 Aachen, Germany
    }



   \abstract
   {Supernova remnants interacting with molecular/atomic clouds are interesting X-ray sources to study broadband nonthermal emission. X-ray line emission in these systems can be produced by different processes, e.g. low energy cosmic rays interacting with the cloud and fast ejecta fragments moving in the cloud.}
   {The paper aims at studying the origin of the non-thermal X-ray emission of the southwestern limb of SN 1006 beyond the main shock, in order to distinguish if the emission is due to low energy cosmic rays diffusing in the cloud or to ejecta knots moving into the cloud.}
   {We analyzed the X-ray emission of the southwestern limb of SN 1006, where the remnant interacts with an atomic cloud, with three different X-ray telescopes (\textit{NuSTAR}, \textit{Chandra} and \textit{XMM-Newton}) and performed a combined spectro-imaging analysis of this region.}
   {The analysis of the non thermal X-ray emission of the southwestern limb of SN 1006, interacting with an atomic cloud, has shown the detection of an extended X-ray source in the atomic cloud, approximately $2$ pc upstream of the shock front. The source is characterized by a hard continuum (described by a power law with photon index $\Gamma\sim1.4$) and by Ne, Si and Fe emission lines. The observed flux suggests that the origin of the X-ray emission is not associated with low energy cosmic rays interacting with the cloud. On the other hand, the spectral properties of the source, together with the detection of an IR counterpart visible with \textit{Spitzer}-MIPS at 24 $\mu$m are in good agreement with expectations for a fast ejecta fragment moving within the atomic cloud.}
   {We detected X-ray and IR emission from a possible ejecta fragment, with radius approximately 1$\times10^{17}$ cm, and mass approximately $10^{-3}M_\odot$ at about 2 pc out of the shell of SN 1006, in the interaction region between the southwestern limb of the remnant and the atomic cloud.}


\titlerunning{Detection of a fast ejecta fragment in SN 1006}
\authorrunning{Giuffrida et al.}

   \maketitle

\section{Introduction}

   \label{sec:intro}

Supernova remnants (SNRs) interacting with interstellar clouds are interesting sources of broadband nonthermal emission. Besides the characteristic hadronic $\gamma-$ray emission associated with $\pi^0$ decay, and the OH maser emission in radio, nonthermal X-rays are also expected. 

Different processes can lead to nonthermal continuum and line emission in X-rays. The bulk of the continuum X-ray emission is typically associated with synchrotron radiation from secondary electrons, the products of $\pi^{\pm}$ decays produced in the interaction of cosmic rays diffusing from the SNR in the nearby cloud, while bremsstrahlung emission from primary and secondary electrons can play an important role in the very hard part of the X-ray band (e.g., \citealt{gac09}).

Nonthermal line emission is also expected. \cite{tdm12} have shown that Low Energy Cosmic Rays (LECRs) interacting with the dense interstellar medium (ISM) can produce the characteristic Fe K$\alpha$ emission line at 6.4 keV, observed in the X-ray spectra of the Arches cluster region, near the Galactic center. This phenomenon can be expected also in supernova remnants interacting with Molecular Clouds (MCs, e. g.,  \citealt{Gabici22}). \cite{nkh19} revealed the presence of two localized regions with enhanced Fe I K$\alpha$ line emission in the northern and central part of IC 443, where the remnant is interacting with extremely dense MCs (e. g., \citealt{cjt22}). This detection was explained as the result of protons at MeV energies accelerated in the SNR and diffusing into the cloud. These particles can eject inner-shell electrons of neutral iron atoms in the cloud, thus producing the K$\alpha$ line emission. A similar scenario can be invoked for the Fe K$\alpha$ emission detected in the region where the SNR W28 is interacting with a MC, as reported by \cite{ohu18} and by \cite{nkm18} (though the two works report enhanced emission in different parts of the remnant). The Fe emission line is consistent with being produced by LECRs, as shown by \cite{pgm20}, who also demonstrated that the enhanced ionization rate in regions near W28 is due to cosmic-ray protons.

Nonthermal X-ray emission in MCs interacting with SNRs can also be observed when fast moving ejecta fragments propagate in the cloud. A theoretical model developed by \cite{b02} (hereafter B02) shows that ejecta knots can produce X-ray nonthermal (continuum and line) emission when interacting with the ISM. The supersonic motion of the ejecta produces a radiative bow shock with prominent infrared  emission. Nonthermal electrons accelerated at the bow-shock diffuse in the fragment, suffering from Coulomb losses and ionizing neutral atoms in the cold clump, thus producing K-shell emission. B02 shows that the line emission increases with the density of the medium, being high when the ejecta knots propagate in MCs. Clear indications strongly supporting this scenario have been obtained by detecting small (albeit extended) hard X-ray emitting sources in IC 443 \citep{bb03,b03, bbp05, zst18} and Kes 69 \citep{bbc12}.

SN 1006 is a Type Ia SNR located well above the galactic plane ($\sim 550$ pc, at a distance of 2.2 kpc, \citealt{wgl03}). Its evolution in a tenuous ISM with density $n_0\sim0.04$ cm$^{-3}$ (\citealt{mbd12, gmc22}) makes it a dynamically young remnant. The shock velocity is of the order of 5000 km s$^{-1}$, though lower velocities are observed in the northwest, and in a local indentation of the southwestern shock most likely due to interaction with a denser environment (\citealt{ksp09, wwb14, mop16}). 
The remnant presents a characteristic nonthermal bilateral emission, with two opposite bright limbs (at northeast and southwest) clearly visible in X-rays, but also in the radio \citep{pdc09}, and $\gamma-$ray band,  \citep{aaa10}. Thermal X-ray emission, mainly associated with ejecta knots, is observed in the northeast, northwest and toward the center.

\begin{figure}
	\includegraphics[width=\columnwidth]{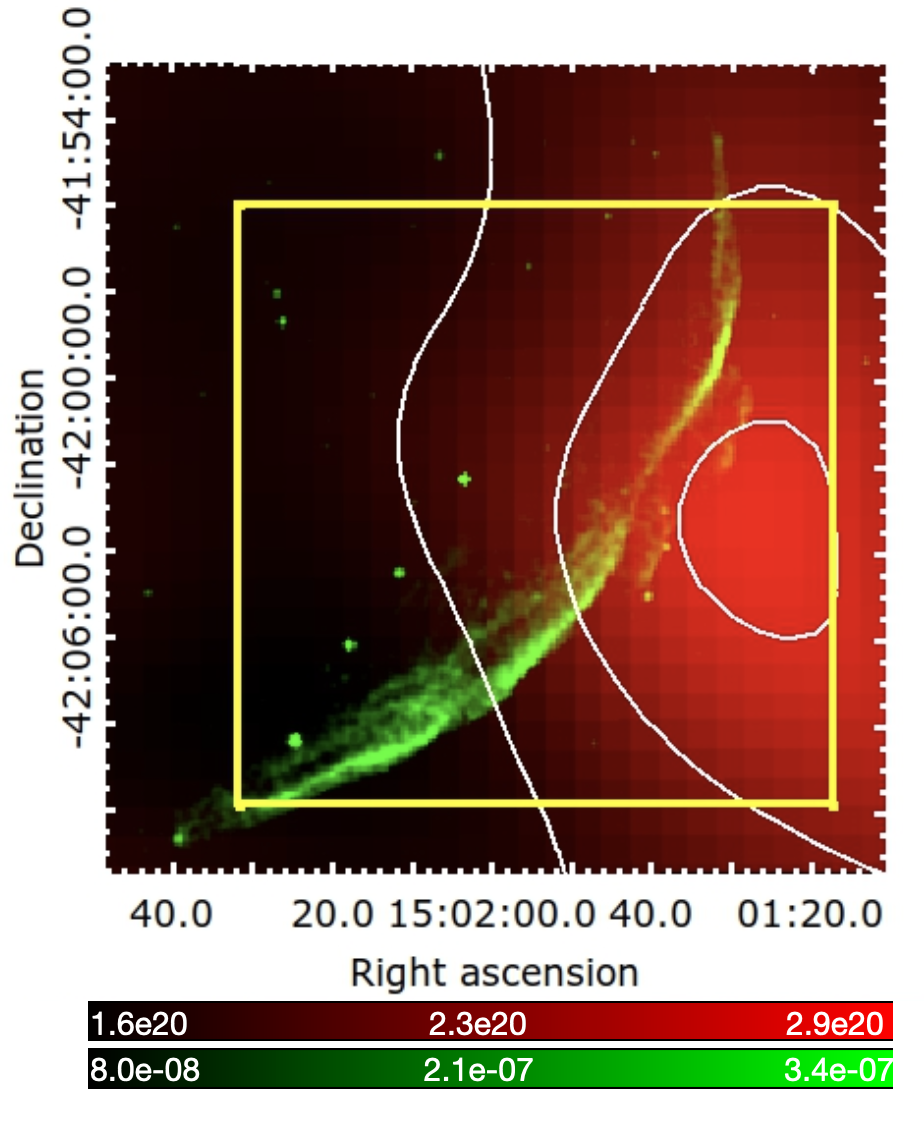}
	\caption{Chandra flux map of SN 1006 in count/s/cm$^2$ in the 2.5-7 keV (green) band. The column density of HI in the $[+5.8,~+10.7]$ km s $^{-1}$ velocity range is shown in red. The contour levels of the column density at the 65\%, 80\% and 95\% of the maximum ($2.9 \times 10^{20}$ cm$^{-2}$)  are shown in white.
 The yellow rectangle shows the \emph{NuSTAR} field of view.}
	\label{fig:rgb_Chandra}
\end{figure}

Despite evolving in a fairly uniform ambient medium, the decrease of the shock velocity shows  that two regions of SN 1006 interact with atomic clouds, namely in the northwest and southwest (see Fig. \ref{fig:rgb_Chandra}). The interaction with an atomic cloud in the northwestern part of the remnant was well studied with multi-wavelength observations (e.g.,  \citealt{lkb88, dgg02, krz04, rks07, abd07, ksl13}) and is associated with a clearly visible H$\alpha$ filament (e. g., \citealt{wwb14}).
The interaction of the southwestern part of SN 1006 with an atomic cloud was studied with radio and X-ray observations \citep{mad14} and modelled with MHD simulations \citep{mop16}.
The combined analysis of X-ray and radio data indicates that the core of the cloud has a density of the order of $n_{core}\sim10$ cm$^{-3}$ (\citealt{mad14}, see also Fig. \ref{fig:rgb_Chandra}). On the other hand, the comparison of the observations with a detailed 3D MHD model, clearly indicates that the part of the cloud actually interacting with the remnant has a density  $n_{cl}\sim0.5$ cm$^{-3}$ \citep{mop16}. 

Recently, a thorough analysis of X-ray observations has shown that SN 1006 can accelerate efficiently both electrons and protons in its nonthermal limbs  \citep{gmc22}. In particular, the efficient hadron acceleration in quasi-parallel conditions (i.e., when the shock normal is almost aligned with the ambient magnetic field) modifies the shock structure, and the shock compression ratio deviates significantly from the canonical value of 4, increasing up to $\sim7$ in the nonthermal limbs (\citealt{mbd12,gmc22}). The inferred azimuthal profile of the compression ratio is in agreement with that expected for modified shocks including the effects of the shock postcursor \citep{hcc20,chc20}.

The southwestern limb of SN 1006 is then characterized by shock-cloud interaction and efficient acceleration of cosmic rays. 
We then expect the southwestern cloud to be a promising source of nonthermal X-rays, which may be associated with LECRs diffusing from the shock to the cloud. On the other hand, nonthermal X-ray emission might be also associated with fast moving ejecta knots decelerating in the cloud. 

This paper presents the analysis of different archival X-ray observations (performed with \emph{NuSTAR}, \emph{XMM-Newton} and \emph{Chandra}) of the southwestern region of SN 1006, where we revealed a small ($\approx 3''$) extended source of nonthermal X-rays (with an infrared counterpart, detected with \emph{Spitzer}), located beyond the shell of the remnant, within the atomic cloud.

The paper is organized as follows: in Sect. \ref{sec:DataRed}  we describe the data reduction procedure; Sect. 3 shows the results of our image and spectral analysis; Sect. \ref{sec:Disc} is dedicated to the discussion on the origin of the nonthermal (and IR) emission, and, finally, in Sect. \ref{sec:conclusion}, we draw our conclusion.

\section{Data reduction}
\label{sec:DataRed}
We here analyzed observations of the southwestern limb of SN 1006 performed with \textit{NuSTAR} (FPMA,FPMB) \citep{lbm18}, \textit{XMM-Newton}/EPIC, \emph{Chandra}/ACIS, and \emph{Spitzer}/MIPS \citep{wgl03}. Table 1 summarizes relevant information about the data.

\begin{table*}
	\caption{List of observations analyzed in this work. All the important information of each observation are included.}
	\begin{center}
		\footnotesize{}
        \resizebox{\textwidth}{!}{
		\begin{tabular}{ccccccccc}
		    \hline\hline
			Telescope & OBS ID & Camera & PI & $\alpha$ & $\delta$ & T$_{exp}$ (ks) & Start Date \\
			\hline
			Nustar & 40110002002 & FPMA/FPMB & J. Li & 15h 01m 49.6s & -42$^\circ$ 02' 47'' & 204.712 & 2016-03-8 \\\\
			\textit{XMM-Newton} & 0653860101 & EPIC-MOS1$^a$ & A. Decourchelle &15h 02m 06.00s  & -42$^\circ$ 05' 26.0'' &   99.578/130.070$^{b}$ & 2010-08-28  \\
		               & & EPIC-MOS2$^a$ & &  & &  101.967/130.070$^{b}$ &\\
			           & & EPIC-pn$^a$ & &  & & 71.623/130.070$^{b}$ & \\\\
            \textit{Chandra (2003)} & 4386 & ACIS-I & Hughes & 15h 02m 07.01s &-42$^\circ$ 07' 30.00'' & 20 & 2003-04-09 \\	
			\textit{Chandra (2012)} & 13738 & ACIS-I & P. F. Winkler & 15h 01m 41.78s & -41$^\circ$ 58' 14.96'' & 73.47 & 2012-04-23  \\
			        & 14424 & ACIS-I & P. F. Winkler & 15h 01m 41.78s & -41$^\circ$ 58' 14.96'' & 25.39 & 2012-04-27  \\\\
			\textit{Spitzer} & 30673$^c$ & MIPS & P. F. Winkler & 15h 01m 54.00s & -41$^\circ$ 53' 00.0'' & 10.383  & 2007-08-22 &  \\
			 & 30673$^d$ & MIPS & P. F. Winkler & 15h 01m 40.00s &-41$^\circ$ 51' 00.0'' & 17.590 & 2007-03-05 &   \\
			\hline
		\end{tabular}}
		\label{tab:data}
	\end{center}
	\tablefoot{$^a$ EPIC-MOS1/MOS2 cameras are in Full Frame Window mode, EPIC-pn camera in Extended Full Frame mode. \\$^b$ Screened/unscreened  exposure time. \\ $^c$ AOR: 18725376 \\ $^d$ AOR: 18725120}
\end{table*}

Data were reprocessed as follows:
\begin{itemize}
    \item \textit{NuSTAR} data analysis was performed with the \textit{NuSTAR Data Analysis Software}, NuSTARDAS, version 1.2.0 with CALDB version 4.9.4 within HEAsoft version v6.28. Data were reprocessed with \texttt{nupipeline}. Spectra were extracted by using the \texttt{nuproduct} pipeline, which also generates the corresponding ancillary response file (\textit{arf}) and redistribution matrix (\textit{rmf}). FPMA and FPMB spectra were fitted simultaneously.
    
    \item \textit{XMM-Newton} data were reprocessed with the \textit{Science Analysis System} (SAS v 19.1.0). 
    We filtered the event files to remove contamination by soft protons with the \texttt{espfilt} task (screened exposure times are shown in Table \ref{tab:data}).
    Images and spectra were produced by selecting events with FLAG=0 and PATTERN $\le4,~12$ for pn and MOS cameras, respectively. Images were background subtracted by adopting the double subtraction procedure described in \cite{mbo17}. For the double subtraction, we used the Filter Wheel Closed (FWC) and Blank Sky (BS) files available at the \emph{XMM-Newton} ESAC repository\footnote{\url{https://www.cosmos.esa.int/web/xmm-newton/filter-closed}\\ \url{http://xmm-tools.cosmos.esa.int/external/xmm_calibration//background/bs_repository/blanksky_all.html}}.
    We produced EPIC mosaicked images by adopting the \texttt{emosaic} SAS task. We corrected the images for vignetting and produced count-rate maps by dividing the superposed EPIC images by the associated superposed exposure maps (obtained with the \texttt{eexpmap} task). The pn exposure maps were multiplied by the ratio of the pn/MOS effective areas, to yield MOS-equivalent superposed count-rate maps. Count-rate maps were then smoothed adaptively throungh the \texttt{asmooth} task.
    Spectra were extracted with the \texttt{evselect} task, while \textit{arf} and \textit{rmf} files were produced with the \textit{arfgen} and \textit{rmfgen} tasks, respectively. We adopted the \texttt{evigweight} task for vignetting correction. 
    
    \item \textit{Chandra} data were analyzed with \textit{CIAO} (v4.13), using \textit{CALDB} (v4.9.4). Data were reprocessed with the \textit{chandra\_repro} task. Flux images of \textit{Chandra} data were obtained by combining the two observation reported in Table \ref{tab:data} with the \textit{CIAO} task \texttt{merge\_obs}.
    
    \item \textit{Spitzer/MIPS} data analysis was performed with the \textit{MOPEX} package (v 18.5.0), which we adopted to produce 24 $\mu$m mosaic images, and to extract point sources from BCD-level data for each observation. An amount of 2114 frames were mosaicked for each observation in 24 $\mu$m band.
\end{itemize}

Spectral analysis was performed using \textit{XSPEC} (v12.11.1d, \citealt{Arnaud1996}).
Spectra include 20 counts per channel for each data set (\textit{XMM-Newton} EPIC-pn/MOS1/MOS2 and \textit{NuSTAR} FPMA/FPMB).

\section{Results}
\label{sec:image}
\subsection{Image analysis} 
\begin{figure*}
    \includegraphics[width=.5\textwidth]{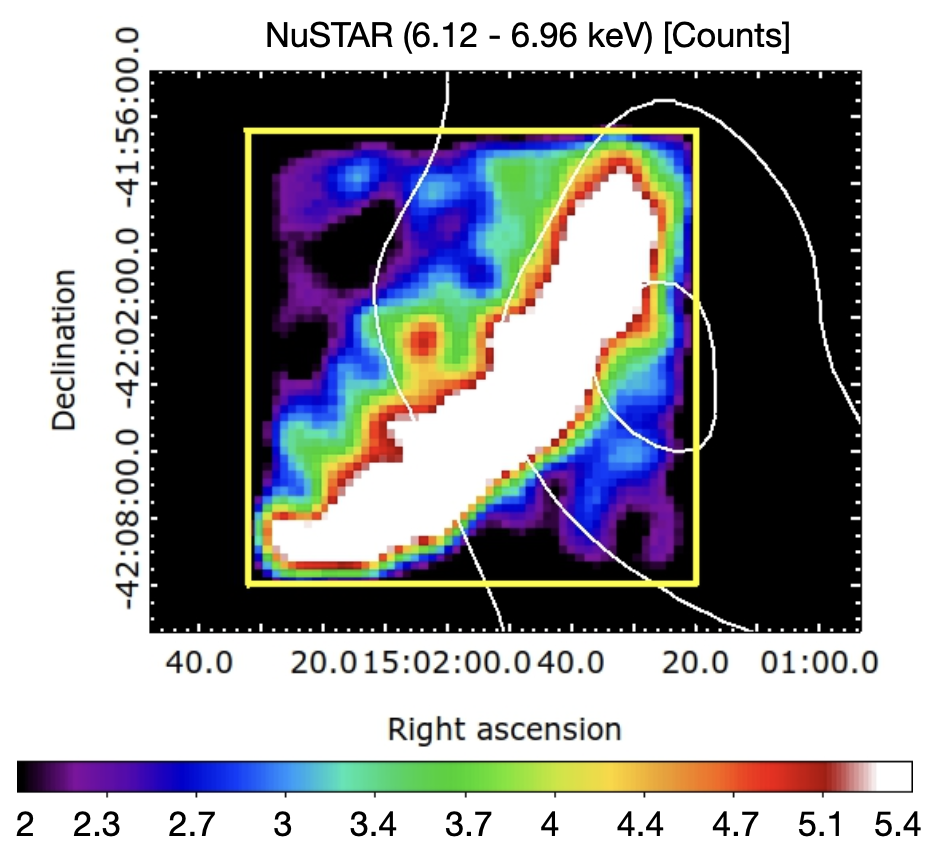}
    \includegraphics[width=.5\textwidth]{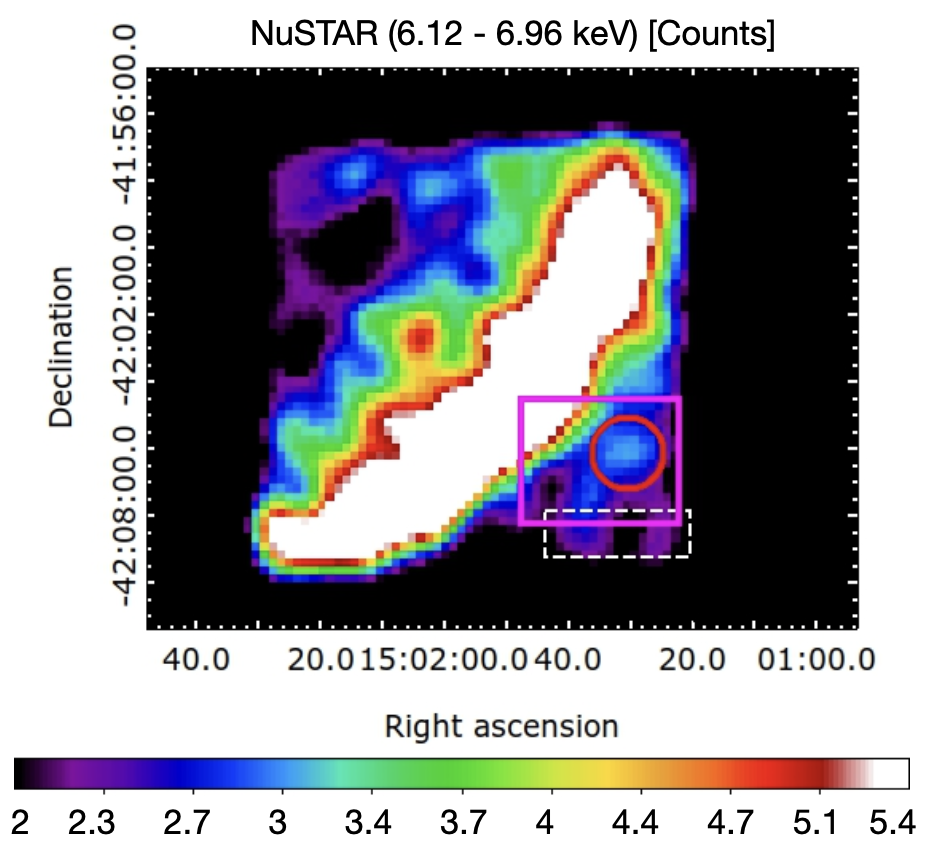}
	\caption{\emph{NuSTAR} observation of the southwestern part of SN 1006 in linear scale obtained as the sum of FPMA and FPMB in the $6.12-6.96$ keV band. The pixel size is $14.7''$. Panel on the left shows (in white) the  contour levels of the column density (Fig. \ref{fig:rgb_Chandra}).
    The red circle on the right shows the region selected for spectral analysis of \textit{knot1} and the white dashed rectangle is the background region}. The magenta box shows the field of view of the three panels in Fig. \ref{fig:xmm-cha_1-7}.
	\label{fig:nustar}
\end{figure*}

In order to constrain the origin of the nonthermal X-ray emission in the interaction region between the southwestern limb of SN 1006 and the atomic cloud we first focused on the Fe K$\alpha$ emission line. Figure \ref{fig:nustar} shows the \emph{NuSTAR} (FPMA and FPMB summed) count image of the southwestern limb of SN~1006 in the $6.12-6.96$ keV band. Beyond the emission from the shell, which is mainly associated with the nonthermal continuum (e.g., \citealt{mbi09}), an isolated knot (which can be called \textit{knot1}) can be spotted outside the forward shock, centered approximately at $\alpha=15\rm{^h}~ 01\rm{^m}~30.4\rm{^s}$, $\delta=-42^\circ~ 06'~ 10.9''$.  The knot is well within the atomic cloud interacting with SN 1006 (see Fig. \ref{fig:nustar}), approximately 2 pc upstream with respect to the shock front (assuming a distance of 2.2 kpc for SN~1006), and its size is comparable to the size of the telescope point-spread function (PSF) of \textit{NuSTAR}.

Motivated by the results obtained with the \textit{NuSTAR} data, we also inspected the \textit{XMM-Newton} and \textit{Chandra} data to exploit the large \emph{XMM-Newton} effective area in the soft X-ray band and the superior spatial resolution provided by \emph{Chandra}. Figure \ref{fig:xmm-cha_1-7} shows the \textit{XMM-Newton} (upper panels) and \textit{Chandra} (lower-left panel) maps of the knot region in the 1-7 keV band. Both in the \textit{XMM-Newton} and \textit{Chandra} maps, we identified a small source{\footnote{The source is 2CXO J150134.1-420620 in the \textit{Chandra} catalog.}, with center coordinates $\alpha=15\rm{^h}~ 01\rm{^m}~ 34.2\rm{^s}$ and $\delta=-42^\circ~ 06'~ 22.8''$ (indicated by a green circle in Fig. \ref{fig:xmm-cha_1-7}), well within the area corresponding to the \emph{NuSTAR} PSF (marked by a red circle in the figure).

To determine whether the source is point-like or extended, we extracted the radial profile of its surface brightness and compared it with simulated PSF data for both \textit{XMM-Newton} and \textit{Chandra} (including \textit{Chandra} observations from both 2003 and 2012). The \textit{Chandra} PSF was obtained with the MARX software (v. 5.5.1), while the \textit{XMM-Newton} PSF was produced with the task \textit{psfgen}. The radial profiles of the source surface brightness are shown in Fig.\ref{fig:psf}. While the poor PSF of \emph{XMM-Newton} does not allow us to resolve the source, for both the 2003 and 2012 \emph{Chandra} data, we observed a significant deviation from the PSF radial profile (i. e., the one expected for point-like sources). We estimate the extension of the source by simulating (with MARX) Gaussian profiles for the source surface brightness, and exploring different values for the sigma, namely $\sigma=2",~\sigma=3",~\sigma=4"$. We find that the observed  profile is well reproduced ($\chi^2=21.4$, with 9 d. o. f.) by the Gaussian with $\sigma=3"$ (see right panel of Fig. \ref{fig:psf}), while the Gaussians with $\sigma=2"$ and $\sigma=4"$ provide a poorer description of the data ($\Delta\chi^2=$6.6 and $\Delta\chi^2=$2.5, respectively)\footnote{We obtain similar results by assuming for the surface brightness a uniform disk with radius $r=3"$, but with a slightly higher value of $\chi^2$ (namely $\chi^2 = 26.1$)}. Our results point toward an extended X-ray source, with a radius of approximately $3''$, corresponding to $\sim10^{17}$ cm at a distance of 2.2 kpc.

The extended X-ray knot which we revealed in the \emph{XMM-Newton} and \emph{Chandra} data may be associated with the \textit{NuSTAR} excess, which we spotted in the Fe K band, considering the PSF of the \emph{NuSTAR} telescope.

We also looked for an optical/IR counterpart of the extended X-ray knot. We did not find any optical counterpart within $15''$ from the center of the \emph{Chandra} source by inspecting the ESO Online Digitized Sky Survey.
On the other hand, we clearly identified the infrared counterpart of the source, detected with \textit{Spitzer} (source ID 2530 in the MOPEX/APEX catalog). Figure \ref{fig:xmm-cha_1-7} (lower-right panel) shows the \emph{Spitzer/}MIPS image (in MJy/sr) at $24~\mu $m, and the source position is indicated by the green circle. The IR flux of the source at 24 $\mu$m is $F_{MIPS} = (6.0 \pm 1.5) \times 10^{-14}$ erg s$^{-1}$ cm$^{-2}$.

\begin{figure*}
	{\includegraphics[width=\textwidth]{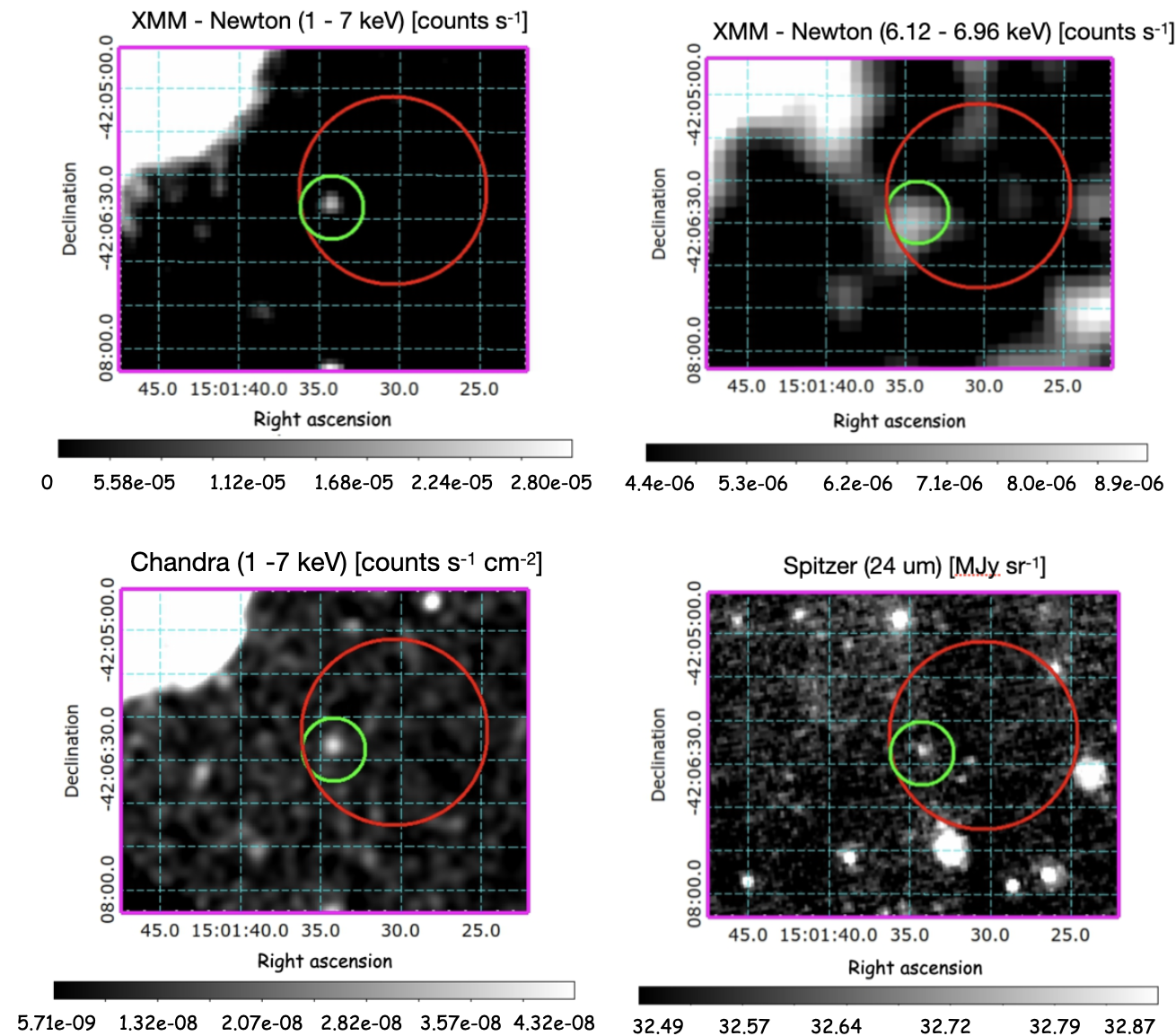}} 
	\caption{Upper left panel: \textit{XMM-Newton}-EPIC count-rate map showing a close-up view of the southwestern limb of SN 1006 in the 1-7 keV band. The bin size is $4''$. The red and green circles indicate the regions adopted to extract the \emph{NuSTAR} and \textit{XMM-Newton} spectra of \textit{knot1}. The field of view corresponds to the magenta box in the right panel of Fig.\ref{fig:nustar}. \emph{Upper right panel:} same as upper left panel in the $6.12-6.96$ keV band, with bin size $8''$. \emph{Lower left panel:} \textit{Chandra} flux image of the same area in the 1-7 keV band (bin size $2''$). \emph{Lower-right panel:}  infrared emission of the same region, as observed with \textit{Spitzer} at 24 $\mu$m.}
	\label{fig:xmm-cha_1-7}
\end{figure*}

\begin{figure*}
\centering
	{\includegraphics[width=.45\textwidth]{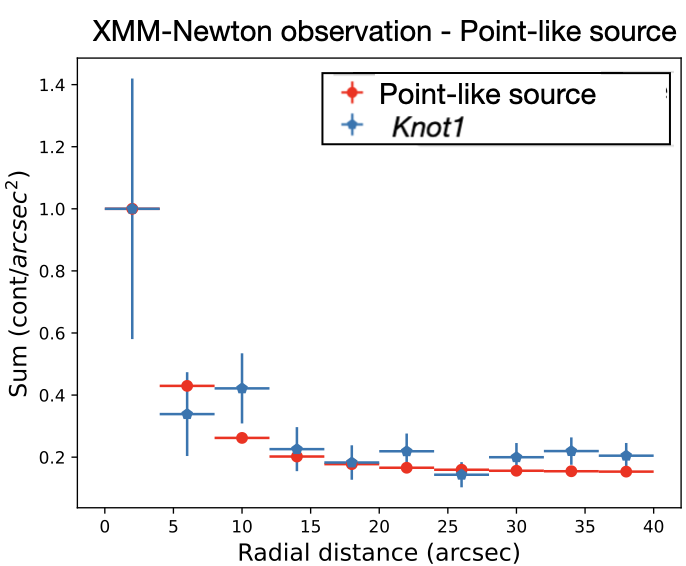}} 
    {\includegraphics[width=.45\textwidth] {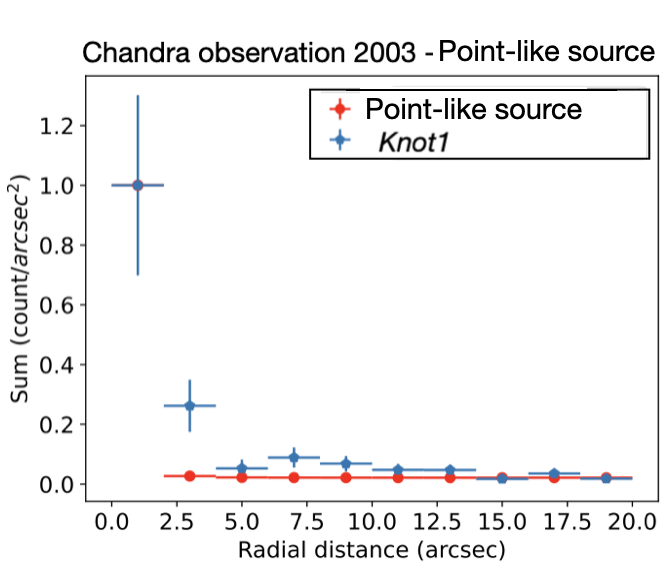}} \\
    {\includegraphics[width=.45\textwidth]{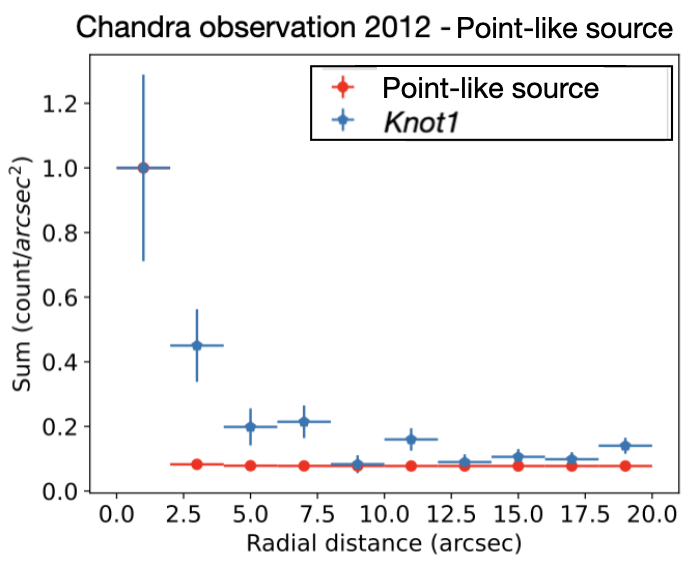}}
    {\includegraphics[width=.45\textwidth]{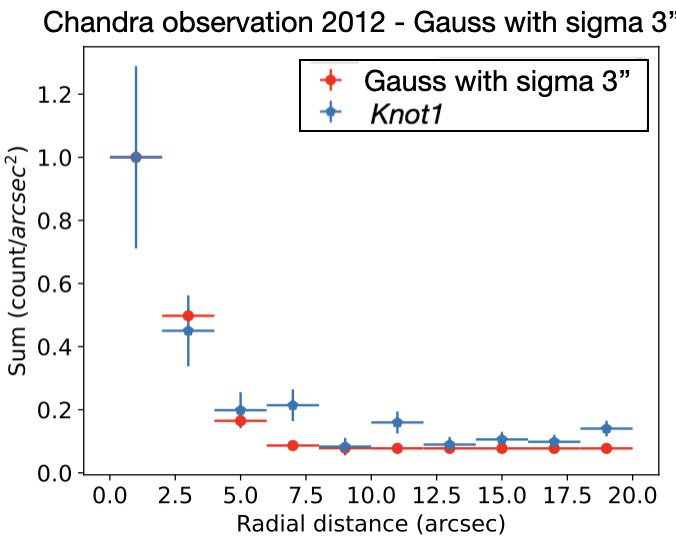}}
	\caption{\emph{Upper-left} panel: comparison between the radial profiles of the \textit{knot1} surface brightness (blue crosses) and that of a simulated point-like source (red crosses, including the contribution of the background) for the \textit{XMM-Newton} data. \emph{Upper-right panel:} same as left panel, but for the \textit{Chandra} 2003 observation. \emph{Lower-left panel:} same as upper-right panel, but for the \textit{Chandra} 2012 observation. \emph{Lower-right panel:} same as lower-left panel, but for a simulated  Gaussian profile with $\sigma=3''$ (red crosses).}
	\label{fig:psf}
\end{figure*}

\subsection{Spectral analysis}
\label{sec:spec}

\begin{table}
	\caption{Best fit parameters for spectral analysis of \textit{knot1}. Errors are at the 68\% confidence level.}
	\begin{center}
		\footnotesize
		\begin{tabular}{c c c}
		    \hline\hline
			Component & Parameter & Best fit value \\
			\hline\\
		    Tbabs & N$_H$ & $8 \times 10^{20}$ cm$^{-2}$ \\\\
		    powerlaw & $\Gamma$ & $1.4_{-0.4}^{+0.5}$\\\\
		    Gauss1 & E & $1.89_{-0.02}^{+0.03}$ keV \\\\
		    Gauss1 & norm ($\times 10^{-7})$ & $2.6_{-0.8}^{+0.8}$ ph cm$^{-2}$ s$^{-1}$ \\\\
		    Gauss2 & E & $6.52_{-0.07}^{+0.08}$ keV\\\\
		    Gauss2 & norm ($\times 10^{-7})$ & $1.9_{-1.0}^{+1.0}$ ph cm$^{-2}$ s$^{-1}$ \\\\
		    Gauss3 & E & $0.89_{-0.02}^{+0.02}$\\\\
		    Gauss3 & norm ($\times 10^{-7})$ & $3.4_{-1.6}^{+1.6}$ ph cm$^{-2}$ s$^{-1}$\\\\
		    \hline
		    & $\chi^2$ & 151.19 \\
		    & d.o.f & 101 \\
			\hline
		\end{tabular}
		\label{tab:fit}
	\end{center}
\end{table}

Spectra were extracted from a circular region with radius $R_{XMM}=22.0''$, $R_{Nu}=65.0''$ and center coordinates $\alpha$= $15\rm{^h}~ 01\rm{^m}~ 34.2\rm{^s}$, $\delta$=$-42^\circ~ 06'~ 22.8''$ and $\alpha$=$15\rm{^h}~ 01\rm{^m}~ 30.4\rm{^s}$, $\delta$=$-42^\circ~ 06'~ 10.9''$ for \emph{XMM-Newton} and \emph{NuSTAR} data, respectively (\emph{Chandra} spectra were not analyzed because of their poor statistics). The background regions are shown in Fig. \ref{fig:nustar}b (white rectangle) and Fig. \ref{fig:XMM_SW} (yellow ellipse) for \textit{NuSTAR} and \textit{XMM-Newton}, respectively.
Spectral analysis was performed on EPIC-pn data in the $0.5-7.5$ keV band, on EPIC-MOS1/MOS2 data in the $0.5-5$ keV band, and on \textit{NuSTAR} data in the $3-10$ keV band. Background spectra were extracted from nearby regions without visible point-like sources. 
\emph{NuSTAR} spectra clearly revealed the presence of an emission line at $\sim6.5$ keV. Figure \ref{fig:spec_nu} shows the FPMA and FPMB spectra in the Fe K spectral band with the corresponding best fit model, consisting in a power law plus one Gaussian component.
On the other hand, Ne and Si emission lines at $\sim0.89$ keV and $\sim1.89$ keV, respectively, can be observed in the \emph{XMM-Newton} spectrum, as shown in Fig. \ref{fig:spec_xmm_Si}.
In the context where the \emph{NuSTAR} knot in the Fe K band and the \emph{Chandra/XMM-Newton} hard extended clump are believed to originate from the same source, we expect that the \emph{NuSTAR} and \emph{XMM-Newton} spectra can be effectively described by the same model simultaneously. We verified that this is indeed the case, and the spectral model consists of a power law plus narrow Gaussian components. We also included the effects of the interstellar absorption (\texttt{tbabs} model within XSPEC) by fixing the interstellar column density at N$_{\rm{H}}=8\times 10^{20}$ cm$^{-2}$ \citep{mad14}, and a multiplicative constant to account for the cross-calibration factor between the two
telescopes, which was left free to vary between 0.9 and 1.1 (in agreement with \citealt{mkc17}) in the fitting process.
Figure \ref{fig:spec_Si+Fe} shows all the spectra with the corresponding best fit models.
Best fit parameters are shown in Table \ref{tab:fit}, errors are at the 68\% confidence level. The global continuum can be accurately represented by a relatively flat power-law function, with photon index $\Gamma=1.4^{+0.5}_{-0.4}$; this is strongly suggestive of a nonthermal origin. We also detected Fe, Si and Ne line complexes with a statistical significance of 95.5\%, 99.99\% and 99.0\%, respectively.

\begin{figure}[ht]
	{\includegraphics[width=\columnwidth]{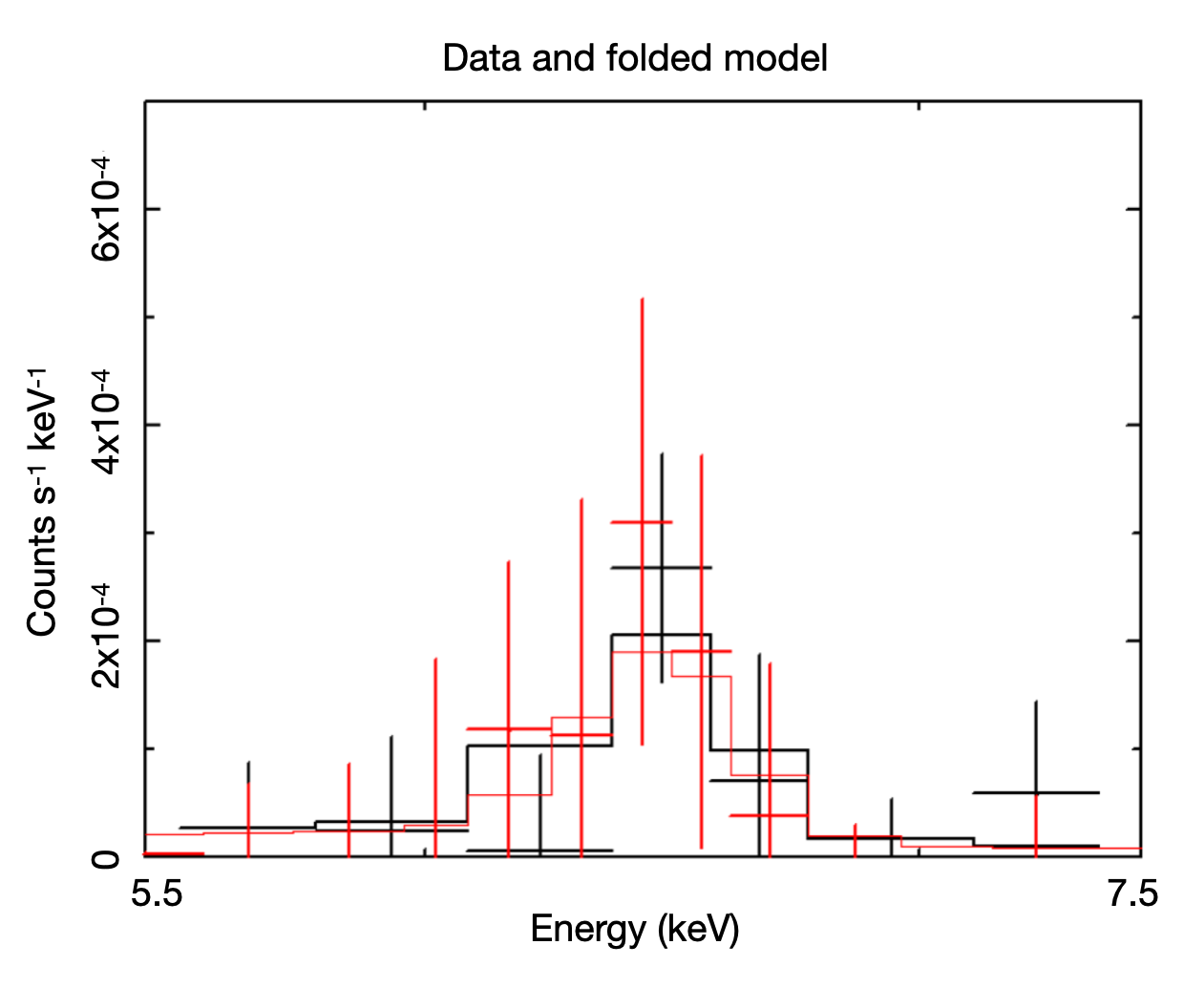}}
	\caption{\textit{NuSTAR} FPMA (black) and FPMB (red) spectra of the \textit{knot1} in 5-7 keV band with the corresponding best fit model.}
	\label{fig:spec_nu}
	{\includegraphics[width=\columnwidth]{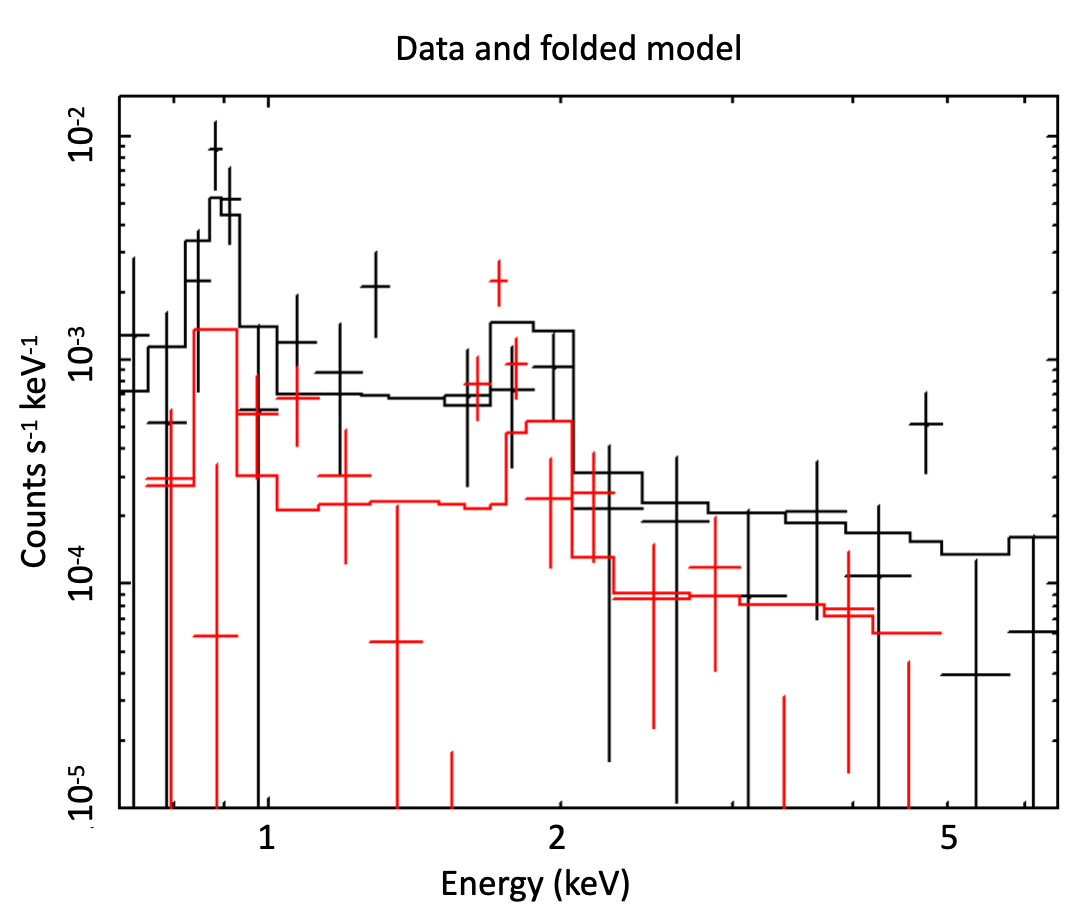}}
	\caption{\textit{XMM-Newton} (EPIC-pn in black, MOS1+MOS2 in red) spectra of the \textit{knot1} with the corresponding best fit model.}
	\label{fig:spec_xmm_Si}
\end{figure}

\begin{figure}
	\includegraphics[width=\columnwidth]{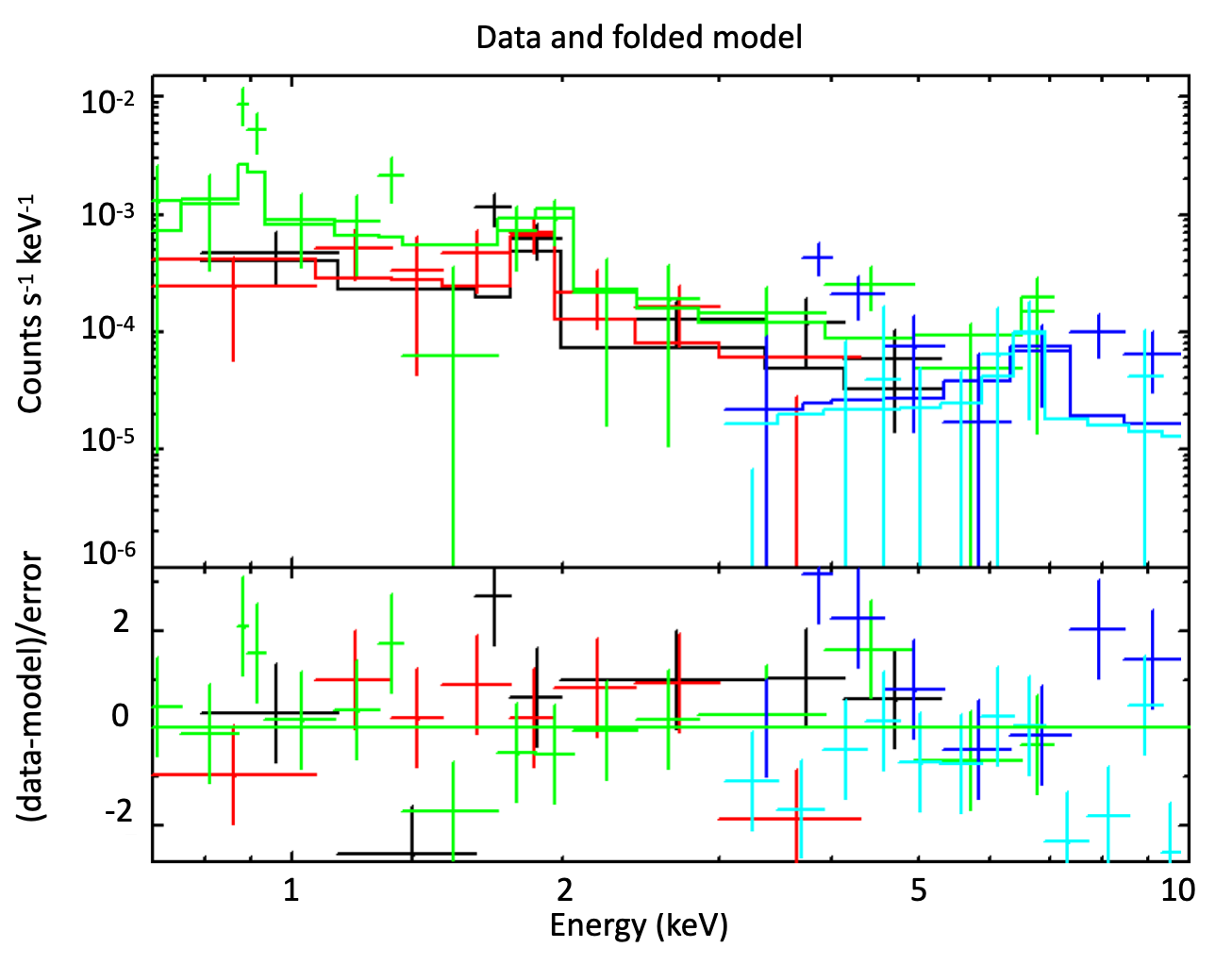}
	\caption{\textit{XMM Newton} (EPIC-MOS1 in black, EPIC-MOS2 in red, EPIC-pn in green) and \textit{NuSTAR} (FPMA in blue, FPMB in light blue) spectra of the \textit{knot1} with the corresponding best fit models and residuals. }
	\label{fig:spec_Si+Fe}
\end{figure}

\section{Discussion}
\label{sec:Disc}
We identified an excess in the Fe emission line (with line centroid at approximately 6.5 keV) in the \emph{NuSTAR} observations of the southwestern region of SN 1006. This excess is consistent with being associated with a small ($\sim 3''$) X-ray knot, visible with \emph{Chandra} and \emph{XMM-Newton}, whose spectrum shows a flat continuum and Si and Ne emission lines. The knot is located at a projected distance of $\sim$ 2 pc from the shell (Fig. \ref{fig:nustar}), where the remnant interacts with an atomic cloud (\citealt{mad14, mop16}) and also shows an infrared counterpart, detected with \emph{Spitzer} at 24 $\mu m$.
The origin of the source can be explained by two different scenarios. We discuss them separately.

\subsection{Diffusion of low energy cosmic rays}
\label{sec:Shock-cloud}
As mentioned in Sect. \ref{sec:intro}, Fe K emission line can be associated with LECRs diffusing from the shock of an SNR to a nearby dense cloud. We then investigated the possibility that particles (electrons and protons), accelerated in the southwestern limb of SN 1006 can produce the observed flux for the Fe line (i. e., $F_{Fe}\sim1.9 \times 10^{-7}$ photons s$^{-1}$ cm$^{-2}$, see Table \ref{tab:fit}) by irradiating the southwestern atomic cloud.
We assumed that LECRs have left the shock front of SN 1006 at the onset of its interaction with the neutral cloud. \citet{mop16} estimated that the shock front reached the atomic cloud $\sim750$ yr after the explosion, hence LECRs have had $t_d\sim250$ yr to diffuse away from the acceleration site. The proton cross section for the production of the Fe K$\alpha$ line peaks at energies $E_p\sim10$ MeV \citep{tdm12}, corresponding to a proton speed $v_p\sim0.14~c$ (where $c$ is the speed of light). We can then set an upper limit for the distance between 10 MeV protons and the acceleration site at the free-streaming value $L=t_d v_p\approx 10$ pc, which is larger than the projected distance between the X-ray emitting knot and the SN 1006 shock front (which is only 2 pc). In principle, it is then possible that LECRs are diffusing within the cloud.
We estimated the expected flux of the Fe K$\alpha$ line associated with cosmic-ray protons and electrons by following the same approach as \cite{pgm20}. The best fit for the multi-wavelength data is shown in Fig. \ref{fig:CRfit}, where we tested different values of the power-law index $\delta$ in the equation for the cosmic-ray density (Eq. \ref{eq:dens_CR_main}). 
\begin{equation}
\label{eq:dens_CR_main}
    n_{CR, i}(p_{i}) = A_i\left(\frac{p_i  c}{\mathrm{MeV}}\right)^{-(\delta+2)}\exp{\left(-\left(\frac{p_i}{p_i^c}\right)^{\beta}\right)}
\end{equation} 
where $i$ corresponds to the species of the CR particle (proton or electron), $A_i$ is the normalisation factor, $p_i$ is the momentum of the particle, $\delta$ is the power-law index and $p_i^c$ is the cut-off momentum. All the details on the model can be found in Appendix A.

\begin{figure}
{\includegraphics[width=\columnwidth]{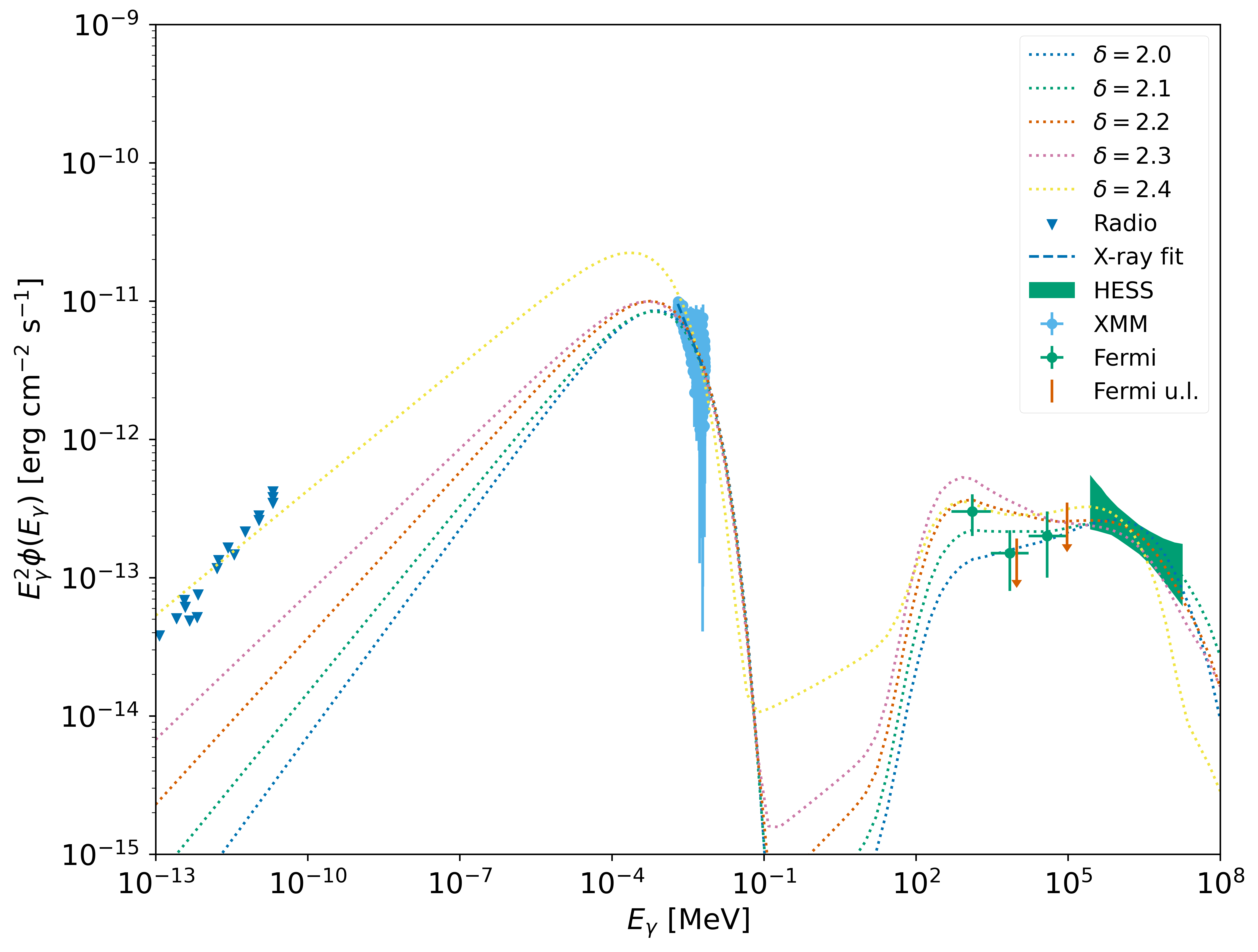}} 
	\caption{Multi-wavelength data fit with emissions from CR interactions for various $\delta$ .}
	\label{fig:CRfit}
\end{figure}

The scenario including $\delta=2.1$ seems to provide the best fit for the multi-wavelength data (Fig.\ref{fig:CRfit}). The Fe K$\alpha$ line intensities expected from the CR protons and electrons are $3.4 \times 10^{-10} \mathrm{cm^{-2}~s^{-1}}$ and $2.9 \times 10^{-11} \mathrm{cm^{-2}~s^{-1}}$ respectively. These values are 3 - 4 orders of magnitude lower than the observed line flux. The mass of the cloud would have to be higher by 3 or 4 orders of magnitude respectively, for the expected flux to match the observed value, which is unlikely, given the location of the remnant at $\sim 550$ pc above the galactic plane. We may hence dismiss the case in which the Fe K$\alpha$ line results from the interaction of LECR with the atomic cloud.

\subsection{Fast ejecta fragment}
\label{sec:Fast_ej_frag}
Another scenario which can explain the X-ray emission of the source shown in Fig. \ref{fig:nustar} and Fig. \ref{fig:xmm-cha_1-7} is based on nonthermal continuum and emission line stemming from fast, metal-rich ejecta fragments in SNRs interacting with a dense ambient medium (B02, \citealt{b03}). A supersonic fragment is preceded by a collisionless bow-shock that enables non-thermal particles to be shock-accelerated. Electrons with keV to MeV energies can diffuse back into the fragment and K-shell ionise neutral matter, resulting in $K{\alpha}$ line emission. By solving a transport equation for non-thermal electrons, \citet{b02} showed that even for conservative choices of the electron acceleration efficiency and diffusion coefficient, this model predicts observable fluxes of $K{\alpha}$ X-rays. Bremsstrahlung from the non-thermal electrons contributes a rather hard continuum with a spectral index of $\approx 1.5$.

Previous works on different remnants, e.g. IC 443 \citep{bb03,bku08} and Kes 69 \citep{bbc12}, have shown an infrared counterpart for X-ray emitting ejecta fragments interacting with interstellar clouds. The X-ray source in analysis shows a significant infrared counterpart at 24 $\mu$m (lower-right panel of Fig. \ref{fig:xmm-cha_1-7}), which is in line with this scenario. 
In particular, as explained in Sect. \ref{sec:DataRed}, we detected a point-like source at $\alpha=15\rm{^h}~ 01\rm{^m}~ 34.1825\rm{^s}~$, $\delta=-42^\circ~ 06'~ 20.447''~$ with a flux density of $6.0\pm0.5 \times 10^{2}$ $\mu$Jy (corresponding to $F_{MIPS}=6.0\pm0.5 \times 10^{-14}$ erg s$^{-1}$ cm$^{-2}$ in the \emph{Spitzer} 24 $\mu$m band, which ranges from $15$ $\mu$m to 30 $\mu$m).  
Fine-structure lines of [FeII] (26 $\mu$m) might provide the main contribution to the flux in the \emph{Spitzer} 24 $\mu$m band. By following \cite{hm89}, we can write the [FeII] (26 $\mu$m) line flux as

\begin{equation}
    F_{26} \approx 2.4 \times 10^{-13} \frac{A}{10^2} \frac{n_{knot}}{10^3} \frac{v_{shk}}{10^2} \; \text{erg} \; \text{cm}^{-2} \; \text{s}^{-1},
    \label{eq:holl_mckee}
\end{equation}
where A is the angular area of the knot in units of arcsec$^2$, $n_{knot}$ is the pre-shock density of the ejecta knot in units of cm$^{-3}$, and v$_{shk}$ is the velocity of the shock moving in the knot in units of km s$^{-1}$. In this scenario, the ejecta knot is moving within the southwestern cloud, so we can consider that $v_{shk}/v_{bow} = \sqrt{n_{cl}/n_{knot}}$ (where $v_{bow}$ is the bow shock velocity and $n_{cl}$ is the cloud density). Therefore, from Eq. \ref{eq:holl_mckee}, we can write
\begin{equation}
    n_{knot} = \left(\frac{F_{26}}{2.4 \times 10^{-13}} \frac{10^2}{A} \frac{10^3}{v_{bow}} \frac{10^2}{\sqrt{n_{cl}}} \right)^2
\end{equation}
where $F_{26}=F_{MIPS}$ 
,
$n_{cl} = 0.5$ cm$^{-3}$ \citep{mop16}, and $A=\pi r^2$ ($r=3"$, as explained in Sec. \ref{sec:image}). assuming that the knot velocity lies in the plane of the sky, we put $v_{bow}$= 6000 km s$^{-1}$, which is obtained by scaling the velocity of the shock in the southwestern limb of SN 1006 (about 5000 km s$^{-1}$, \citealt{wwb14}) by the factor $f=d_{knot}/R_{1006}$, where $R_{1006}$ and $d_{knot}$ are the distances of the shock front and of the knot, respectively, from the center of the remnant.We then get $n_{knot}= (431 \pm 64)$ cm$^{-3}$, corresponding to a mass $M_{knot}=(1.4 \pm 0.2) \times 10^{-3}$ M$_\odot$, assuming solar abundances. 
We caution the reader that Eq. \ref{eq:holl_mckee}, 3  were derived for a knot with solar abundances \citep{hm89}. Since the chemical composition strongly affects the efficiency of radiative cooling, and then the temperature and density profile of the shock, a proper correction of Eq. \ref{eq:holl_mckee}, 3 for a pure-metal knot is hard to evaluate.
In this case, we need to assume that the mass of the knot is of the order of $10^{-3} M_\odot$.

The parameters derived for the X-ray and IR emitting knot are similar to those considered in B02, who modelled the emission stemming from a fast moving ejecta fragment with radius R $\sim3 \times 10^{16}$ cm and mass M $\sim10^{-3}M_{\odot}$. In that case the knot is composed predominantly by oxygen, with $\sim 10^{-4}$ M$_\odot$ of impurities (Si, S, Ar, Ca, Fe). B02 analyzed two different scenarios, namely i) ejecta fragments moving in  a dense ($10^3$ cm$^{-3}$) molecular cloud, and ii) in a low density medium ($0.1$ cm$^{-3}$). 

In order to compare the emission predicted by B02 with that observed by us, we focused on the second scenario, where the density is similar to the density of the atomic cloud interacting with the southwestern part of SN 1006 ($n_{cl}\sim0.5$ cm$^{-3}$, \citealt{mop16}). Indeed, according to B02, the X-ray flux is expected to increase with the ambient density, so the values predicted by B02 should be considered as lower limits for the case of SN 1006, where the ambient density is a factor of 5 higher. 

Table \ref{tab:byk} shows the comparison between the X-ray emission properties predicted by B02 and the results obtained with our spectral analysis of the knot interacting with the atomic cloud in the southwestern part of SN 1006. We found that the continuum emission is in good agreement with expectations. In particular, the photon index is $<1.5$ (as predicted in B02) and the X-ray luminosity  (in the 4-10 keV band) is $>10^{30}$ erg/s, in agreement with expectations for knots larger than $10^{17}$ cm (see B02 for details). On the other hand, we observed higher luminosity for the Si and Fe line complexes than those predicted by B02 for an O-rich knot. In particular, the Si-line luminosity and the Fe-line luminosity are about 20 and 50 times more than the expected values, respectively. This effect can be due to differences in the chemical composition of the ejecta knot since the knot in B02 contains only 10\% of Si, S, Ar, Ca and Fe. In the case of \textit{knot1}, the chemical composition suggests that it originates from the innermost part of the remnant (high velocity Fe rich knots have been observed in type Ia supernovae, \citealt{rhg14}). 

We have assumed the \textit{NuSTAR} source to be associated with the same source observed in the \textit{XMM-Newton}, \textit{Chandra} and \textit{Spitzer} data. Nevertheless, our conclusions stay unaffected even if we remove this assumption and do not include the Fe line in the spectral fittings. In particular, we recovered the same flux and photon index as that obtained.

\subsection{Proper motion}

\begin{figure*}[ht!]
	{\includegraphics[width=\textwidth]{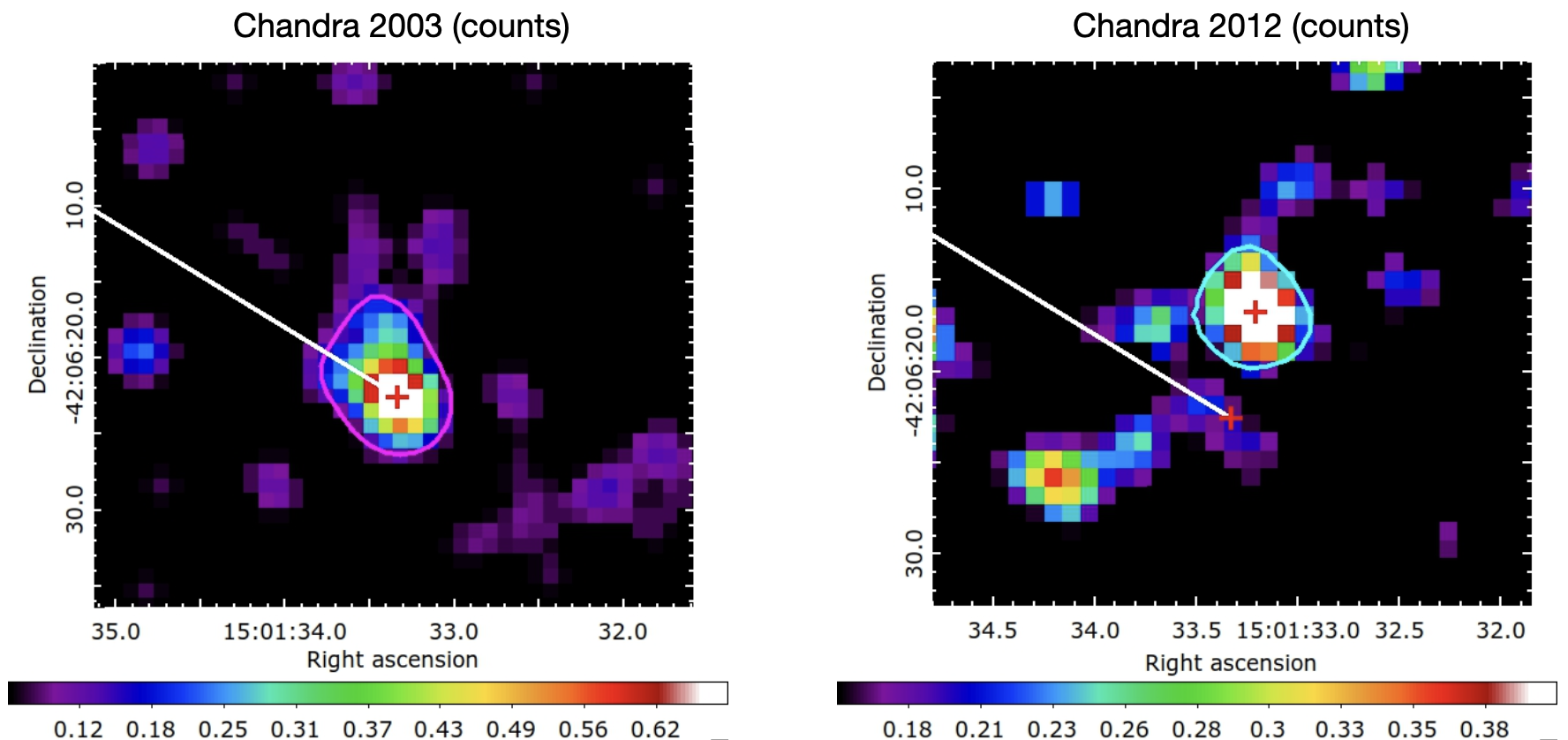}}
 	{\includegraphics[width=\textwidth]{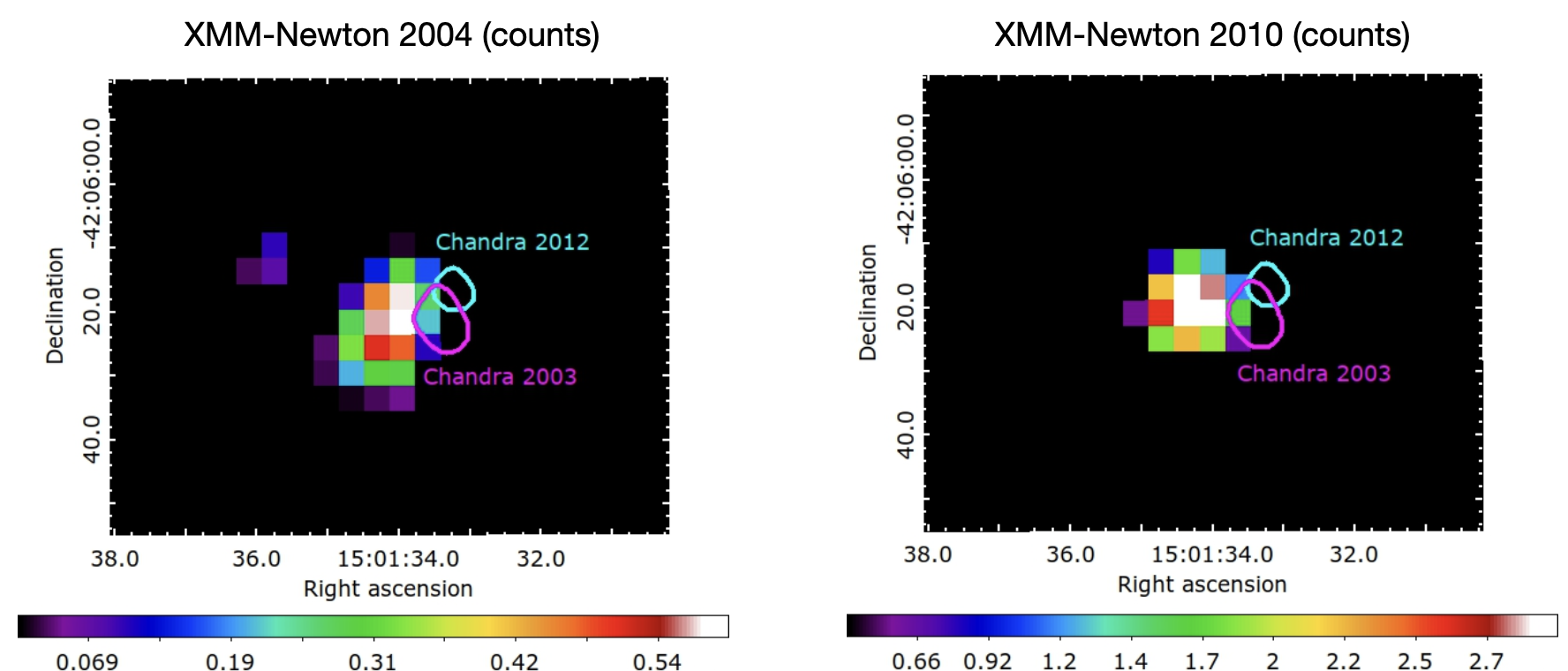}}
	\caption{Upper left panel: \textit{Chandra} counts image observed in 2003. The red point shows the position of the source and the white line indicates the direction to the center of the remnant. The magenta region marks the shape of the source. Upper right panel: \textit{Chandra} counts image observed in 2012. The northern red point indicates the position of the source and he cyan region marks its shape.
 Lower left panel: \textit{XMM-Newton} counts image observed in 2004. Lower right panel: \textit{XMM-Newton} counts image observed in 2010. The magenta and the cyan regions in both the lower panels indicate the source detected with \textit{Chandra} in 2003 and 2012 respectively.}
	\label{fig:pm}
\end{figure*}

In order to verify the effective motion of the knot we checked its proper motion using \textit{Chandra} observations. We compared the data collected in 2012 (ObsID 13738 and 14424) and in 2003 (ObsID 4387) using an absolute coordinate system as in \cite{wwb14}. We then run the tool \texttt{wavedetect} to find X-ray point-like sources in each observation and we run \texttt{wcs match} to match the X-ray sources with their optical counterpart detected in the NOMAD catalog. Unfortunately, the limited number of counts available in both observations hamper the possibility of performing a robust analysis. Results are shown in Fig.\ref{fig:pm}. The knot seems to move from an initial position (2003, Fig.\ref{fig:pm} upper-left panel), to a final position (2012, Fig.\ref{fig:pm} lower-right panel) of 5 arcsec in the north-west direction.
However, this is not confirmed by the 2004 and 2010 \textit{XMM-Newton} observations, where no significant proper motion is detected (lower panels of Fig. \ref{fig:pm}). For the \textit{XMM-Newton} analysis, we used the combined count-rate image taking into account pn, MOS1, MOS2.
A new \emph{Chandra} observation is necessary to provide a more reliable assessment of the proper motion of the source.

\subsection{Alternative interpretations}
We cannot exclude the possibility that the source is not associated with SN 1006. The X-ray emission from \textit{knot1} could in principle have a galactic or an extragalactic origin. The significant extension of \textit{knot1} clearly points against a stellar origin or a compact object, thus making the galactic scenario unlikely.
The Fe-K$\alpha$ line and a flat continuum below 10 keV ($\Gamma$<2) are indeed among the main characteristics of AGNs and Seyfert galaxies. Moreover, it has been found that some of them can show extended X-ray emission, such as in \citet{arevalo,bauer,Fabbiano17,fabbiano18a,fabbiano18b,fabbiano19,Makysm17,jones20,Travascio21}. However, an optical counterpart should be visible, which is not the case for \emph{knot1}, and no proper motion should be visible. The extragalactic catalogs do not provide any detection located at \textit{knot1}. 

\begin{table}
	\caption{Comparison between results found in this paper and predictions by B02.}
	\begin{center}
		\footnotesize
		\begin{tabular}{c c c}
		    \hline\hline
			Parameter & This paper & B02 \\
			\hline\\
		    Radius (cm) & 1$ \times 10^{17}$ & $3 \times 10^{16}$  \\\\
		    Mass (M$_\odot$) & $(1.4 \pm 0.2) \times 10^{-3}$ &  10$^{-3}$\\\\
		    Ambient density (cm$^{-3}$) & 0.5 & 0.1\\\\
		    Si luminosity (ph s$^{-1}$) & $1.6 \times 10^{38}$ & $6.7 \times 10^{36}$ \\\\
		    Fe luminosity (ph s$^{-1}$) & $1.0 \times 10^{38}$ & $2.1 \times 10^{36}$\\\\
		    L$_{X}$ ($4-10$ keV) (erg s$^{-1}$) & $3.5 \times 10^{30}$ & $\gtrsim 10^{30}$ $^{(a)}$\\\\
		    Photon index & $1.4_{-0.4}^{+0.5}$ & $\leq$ 1.5\\\\
			\hline
		\end{tabular}
		\label{tab:byk}
	\end{center}
	\tablefoot{We calculated all of the luminosity from fluxes in Table \ref{tab:fit} assuming a distance of 2.2 kpc. \\$^{(a)}$ Calculated for fragment larger than $10^{17}$ cm (B02).}
\end{table}

\section{Summary and conclusion}
\label{sec:conclusion}
We analyzed the X-ray emission of the southwestern limb of SN 1006, where the remnant is interacting with an atomic cloud (\citealt{mad14, mop16}), with three different X-ray telescopes (\textit{NuSTAR}, \textit{XMM-Newton} and \textit{Chandra}). 

We discovered an X-ray knot out of the shell (about 2 pc upstream of the shock front), which is clearly visible in the \emph{XMM-Newton} and \emph{Chandra} data and shows an IR counterpart, which we observed in the \emph{Spitzer} MIPS 24 $\mu$m data.
In a region compatible with the \textit{Chandra} and \textit{XMM-Newton} localization, the analysis of \textit{NuSTAR} data indicates the presence of an X-ray source within the atomic cloud interacting with the southwestern limb of SN 1006 (Fig. \ref{fig:nustar}) whose size is comparable with the PSF of the telescope. The combined analysis of \textit{XMM-Newton} and \textit{Chandra} observations, with their higher effective area and spatial resolution, allowed us to constrain the location and the extension of the source. As a result, we found a knot centered at $\alpha=15\rm{^h}~ 01\rm{^m}~ 34.2\rm{^s}$ and $\delta=-42^\circ~ 06'~ 22.8''$ with radius $R \sim$ $1 \times 10^{17}$ cm (assuming the same distance as SN 1006).

Spectral analysis of the X-ray knot shows three significant emission lines at 0.89 1.89 keV and 6.5 keV (see Table \ref{tab:fit} and Fig. \ref{fig:spec_Si+Fe}) associated with Ne, Si and Fe, respectively. For their origin we have considered two different scenarios:

\begin{enumerate}
    \item Low energy cosmic rays diffusing from the shock in the SW limb of SN 1006 to the atomic cloud produce non-thermal emission lines, especially the characteristic Fe-K$\alpha$ line at 6.4 keV. However the CR spectra that best fit the multi-wavelength observations of the SW limb produce a Fe-K$\alpha$ line intensity that is several orders of magnitude lower than the observed intensity.
    
    \item Fast ejecta fragments in SNR interacting with interstellar clouds produce infrared emission and  nonthermal X-ray emission, characterized by a hard continuum and emission lines. The presence of an IR counterpart for the isolated knot in SN 1006, together with its X-ray flux and spectral shape, are in nice agreement with the predictions by B02. We report higher luminosities for the emission lines than those predicted by B02 and interpret this as the result of a different chemical composition of the ejecta knot. We estimated the physical parameters of the ejecta knot, finding a density $n_{knot}= (431 \pm 64)$ cm$^{-3}$ and a mass M$_{knot} = (1.4 \pm 0.2) \times 10^{-3} M_{\odot}$. 
\end{enumerate}

Nonthermal emission from fast ejecta knots has been observed only in the core-collapse SNRs IC 443 \citep{bku08} and Kes 69 \citep{bbc12}. This paper provides the first indication of an Ne/Si/Fe-rich fragment of ejecta in a Type Ia SNR. The proper motion is crucial to confirm that the \textit{knot1} is a fast ejecta knot associated with SN 1006.

\section*{Acknowledgments}
This work was partially supported by the INAF mini-grant ``X-raying shock modification in supernova remnants",
VHMP acknowledges support from the Initiative Physique des Infinis (IPI), a research training program of the Idex SUPER at Sorbonne Universit\'e.

This study was also partially supported by the LabEx UnivEarthS, ANR-10-LABX-0023 and ANR-18-IDEX-0001.

\appendix

\section{Model for LECRs diffusing in the cloud}

\begin{table*}[ht!]
\footnotesize
	\caption{Multi-wavelength fit parameters normalisation $A_i$ ($\mathrm{MeV^{-3}~{cm^{-3}}}$) and momentum cut-off $p_i^c$ ($\mathrm{MeV}$) for various CR spectral indices $\delta$ and corresponding Fe K$\alpha$ line intensities (photons $\mathrm{cm^{-2} s^{-1}}$).}
	\begin{center}
		\begin{tabular}{c c c c c c}
		    \hline\hline
			$\delta$ & CR species & $A_i$ & $p_i^c$ & $I_{K{\alpha}}^i$ \\
			\hline\\
		    2.0 & p / e$^-$& $2 \times 10^{-5}$ / $10^{-9}$ & $2 \times 10^8$ / $2 \times 10^7$ & $1.3 \times 10^{-10}$ / $5.1 \times 10^{-12}$ \\\\
            2.1 & p / e$^-$ & $9 \times 10^{-5}$ / $5 \times 10^{-9}$ & $5 \times 10^8$ / $2 \times 10^7$  & $3.4 \times 10^{-10}$ / $2.9 \times 10^{-11}$    \\\\
		    2.2 & p / e$^-$ & $4 \times 10^{-4}$ / $3 \times 10^{-8}$ & $5 \times 10^8$ / $2 \times 10^7$ & $8.7 \times 10^{-10}$ / $2.0 \times 10^{-10}$  \\\\
            2.3 & p / e$^-$ & $1.5 \times 10^{-3}$ / $1.5 \times 10^{-7}$ & $10^9$ / $2 \times 10^7$ & $1.9 \times 10^{-9}$ / $1.1 \times 10^{-9}$ \\\\
		    2.4 & p / e$^-$ & $2 \times 10^{-3}$ / $2 \times 10^{-6}$ & $10^9$ / $1.5 \times 10^7$  & $1.5 \times 10^{-9}$ / $1.7 \times 10^{-8}$      \\\\
            2.5 & p / e$^-$ & $7 \times 10^{-3}$ / $5 \times 10^{-6}$ & $10^9$ / $2 \times 10^7$ & $3.1 \times 10^{-9}$ / $4.9 \times 10^{-8}$   \\\\
		    2.6 & p / e$^-$ & $1 \times 10^{-2}$ / $2.5 \times 10^{-5}$ & $10^9$ / $2 \times 10^7$ & $2.7 \times 10^{-9}$ / $2.9 \times 10^{-7}$  \\\\
            2.8 & p / e$^-$ & $3 \times 10^{-2}$ / $7 \times 10^{-4}$ & $10^9$ / $2 \times 10^7$  & $2.9 \times 10^{-9}$ / $1.1 \times 10^{-5}$ \\\\
			\hline
		\end{tabular}
		\label{tab:CRfit}
	\end{center}
\end{table*}

In principal, fitting the non-thermal emissions from both the shock in the SW limb and the cloud requires a propagation model to describe the escape of CRs from the shock regions into the cloud. Such a model, however, necessitate quite a few free parameters, e.g. the shape and normalization of the diffusion coefficient of CRs in these regions over a broad energy range from MeV to TeV. In order not to inflate the number of free parameters, we will adopt a more simplified approach which is to assume that the CR density is uniform both in the shock and in the cloud regions. This assumption will lead to a conservative upper limit for our predictions of the Fe K$\alpha$ emission induced by low-energy CRs escaping from the shock. We will see later that this upper limit is expected to be much lower than the observed Fe K$\alpha$ emission and, thus, more complicated propagation model might not be required.

We estimated the expected flux of the Fe K$\alpha$ line associated with cosmic rays protons and electrons by following the same approach as \cite{pgm20}. In all following equations, the mass and momenta of the particles are given in $\mathrm{MeV}$ meaning $m_ic^2$ and $p_ic$ for a particle $i$. In particular, we assume a CR density of the form :
\begin{equation}
\label{eq:dens_CR}
    n_{CR, i}(p_{i}) = A_i\left(\frac{p_i}{\mathrm{MeV}}\right)^{-(\delta+2)}\exp{\left(-\left(\frac{p_i}{p_i^c}\right)^{\beta}\right)}
\end{equation} 
where $i$ corresponds to the species of the CR particle (proton or electron), $v_i$ is the particle speed, $A_i$ is the normalisation factor ($\mathrm{MeV^{-3}~cm^{-3}}$), $p_i$ is the momentum of the particle, $\delta$ is the power-law index and $p_i^c$ is the cut-off momentum. The exponential $\beta$ of the cut-off is equal to $1.0$ for protons and $2.0$ for electrons as suggested by \cite{Zira} for electrons accelerated in shell-type supernova remnants. The parameters $\delta$, $A_i$ and $p_i^c$ can be determined by fitting the available multi-wavelength data:
\begin{itemize}
    \item Radio data from the entire remnant (\citealt{Reynolds})
    \item \emph{XMM-Newton} X-ray data from the SW limb (see Appendix B)
    \item $\gamma$-ray data from FERMI and \emph{HESS} both from the SW limb (\citealt{Xing}, \citealt{Acero15}, \citealt{Acero})
\end{itemize}

Additionally, we fitted the \emph{XMM-Newton}/EPIC-pn spectrum of the southwestern limb of SN 1006 between 2 $\mathrm{keV}$ and 6 $\mathrm{keV}$ as 
\begin{equation}
    F_{X-ray}(E) = 1.14\times10^{-2}\biggr(\frac{E}{keV}\biggr)^{-2.938}\mathrm{cm^{-2}~s^{-1}~keV^{-1}}
\end{equation}

The main contribution from CR protons is the $\gamma$-ray emission. Following \cite{Kafexhiu} the spectrum of $\gamma$ rays $\Phi_\gamma$ ($\mathrm{MeV^{-1}~cm^{-3}~s^{-1}}$) resulting from the p-p interactions of a proton intensity $J_{p}$ ($\mathrm{MeV^{-1}~cm^{-2}~s^{-1}~sr^{-1}}$)can be expressed in the following way:

\begin{equation}
    \Phi_\gamma(E_\gamma) = 4\pi{n_H}\int\frac{d\sigma}{dE_\gamma}(E_p, E_\gamma)J_p(E_p)\,dE_p
\end{equation}

We modelled leptonic $\gamma$-ray emission via inverse Compton (IC) scattering and relativistic Bremsstrahlung following \cite{Cristofari}. 
The spectrum of IC $\gamma$-rays $\Phi_{IC}$ ($\mathrm{MeV^{-1}~cm^{-3}~s^{-1}}$) resulting from an electron density $N_e$ ($\mathrm{MeV^{-1}~cm^{-3}}$) and a seed photon field (T, $\kappa$) where $T$ is the temperature and $\kappa$ is the dilution factor, is expressed in the following way (\citealt{Khangulyan}):
\begin{equation}
    \Phi_{IC}(E_\gamma) = \int\frac{dN_{iso}}{d\omega{dt}}(E_\gamma, E_e, T, \kappa)N_e(E_e)\,dE_e
\end{equation} 

The spectrum of $\gamma$-rays resulting from relativistic Bremsstrahlung $\Phi_{Brem}$ ($\mathrm{MeV^{-1}~cm^{-3}~s^{-1}}$) produced by an electron density $N_e$ ($\mathrm{MeV^{-1}~cm^{-3}}$) is as follows (\citealt{Dogiel}):
\begin{equation}
    \Phi_{Brem}(E_\gamma) = c{n_H}\int\sigma_{Brem}(E_\gamma, E_e)N_e(E_e)\,dE_e
\end{equation}
where $\sigma_{Brem}$ is the related cross-section from \cite{Schlickeiser}.

CR electrons also contribute to the radio and X-ray domains through synchrotron emission. The spectrum of synchrotron emission $\Phi_{syn}$ ($\mathrm{MeV^{-1}~cm^{-3}~s^{-1}}$) from an electron density $N_e$ ($\mathrm{MeV^{-1}~cm^{-3}}$) in a magnetic field $B_0$ can be expressed in the following way (\citealt{Celli}):
\begin{equation}
    \Phi_{syn}(E_\gamma) = \frac{\sqrt{3}{e^3}B_0}{2\pi{m_e}{\hbar}{E_\gamma}}\int_0^{\infty}R\biggr(\frac{E}{E_c(E_e)}\biggr)N_e(E_e)\,dE_e
\end{equation} where $e$ is the elementary charge, $m_e$ is the rest mass energy of the electron and where $E_c = 0.07\left(\frac{B_0}{mG}\right)\left(\frac{E_e}{TeV}\right)^2 \mathrm{keV}$, where $B_0$ is the magentic field and $E_e$ the kinetic energy of the electron. The function $R$ from \citealt{Zira10} is defined as:
\begin{equation}
    R(x) = \frac{1.81e^{-x}}{\sqrt{x^{-2/3} + (3.62/\pi)^2}}
\end{equation}

We calculated the $\gamma$-ray emission from protons and electrons and synchrotron emission from electrons from the southwestern shell and cloud. According to \cite{mop16}, we assumed a spherical cloud with radius $R_{cl} \approx 5.5\times10^{18}~ \mathrm{cm}$ and mass $M_{cl} = 0.4~\mathrm{M}_\odot$. The average gas density $n_{cl}$ is $0.5~ \mathrm{cm}^{-3}$ (\citealt{mop16}). The magnetic field $B_{cl}$ is not precisely known but expected to be of the order of a few micro Gauss. We take $B_{cl}$ to be $10 ~\mu\mathrm{G}$ to maximise synchrotron emission. Following \cite{Cristofari}, the radius $R_{sh}$ is taken as $7.7 ~\mathrm{pc}$ and the volume-filling factor $\xi_f$ is $0.25$ as we only consider the SW limb. The volume of this part of the shell is then $V_{sh} = (\xi_f/3)\pi{R_{sh}^3}$. The post-shock gas density $n_{sh}$ is $0.16 ~\mathrm{cm}^{-3}$ (\citealt{gmc22}). The magnetic field $B_{sh}$ is $90~ \mu\mathrm{G}$ (\cite{wwb14}). 

We tried fitting the $\gamma$-ray and X-ray data using various different CR spectral indices $\delta$ between $2.0$ and $2.8$. The radio data points were taken as upper limits as they were obtained from the entire remnant. Using the best fit parameters for each spectral index, we computed the expected Fe K$\alpha$ line intensities from all the different CR spectra. 

The intensity of the Fe K$\alpha$ line resulting from CR interactions can be expressed as:
\begin{equation}
    I_{K{\alpha}}^i = \frac{M_{cl}}{4\pi{d_s^2}{m_{avg}}}\int_{T_i^{th}}^\infty\sigma_{K{\alpha}}^i(E_i)v_iN_i(E_i)\,dE_i
    \end{equation}
where $i$ represents the species of the CR particle, $\sigma_{K{\alpha}}^i$ is the K-shell ionisation cross-section by CR species $i$ (\citealt{Tatischeff}) which takes into account solar abundance of Fe ($\eta_{Fe}=3 \times 10^{-5}$) and $m_{avg}=1.4m_H$. The intensity was obtained assuming a minimum ionising energy of $10$ keV. 

The expected intensities from the CR spectra are a few orders of magnitude lower than the measured Fe K$\alpha$ line intensity. We also estimated the necessary cloud mass to have the measured intensity from these spectra. These values $I_{K{\alpha}}^i$ (photons $\mathrm{cm^{-2}.s^{-1}}$) and expected $M_{cl}$ ($M_\odot$) are given in Table \ref{tab:CRfit}.

\section{\textit{XMM-Newton} observations in the SW limb}
Fig. \ref{fig:XMM_SW} shows the \textit{XMM-Newton} flux image in counts/s of the southwestern limb of SN 1006. The region in black has been used to extract the spectrum useful to fit the X-ray data in Sec. \ref{sec:Shock-cloud}.
We also explored different background regions and verified that the best fit values do not depend on that.

\begin{figure}[h]	
{\includegraphics[width=\columnwidth]{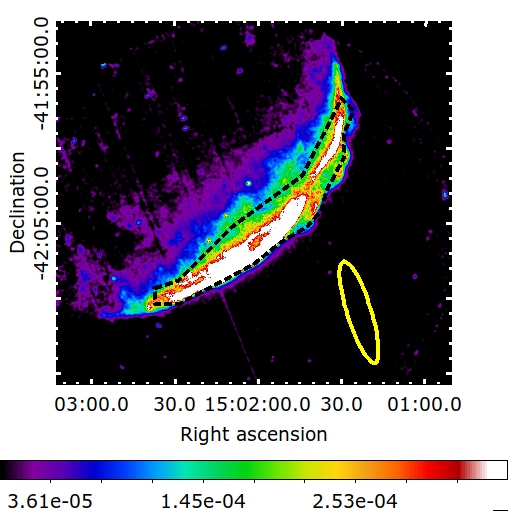}}
	\caption{\textit{XMM-Newton} flux image in counts/s of the SW limb of SN 1006. The black-dashed line shows the region useful for the X-ray data analysis from the SW limb. The yellow region marks the background region used to extract the EPIC-pn spectrum.}
	\label{fig:XMM_SW}
\end{figure}

\bibliographystyle{aa}
\bibliography{references}

\end{document}